\documentclass[twocolumn,english,aps,prl,showpacs]{revtex4}
\usepackage[T1]{fontenc}
\usepackage[latin1]{inputenc}
\usepackage{amsmath}
\usepackage{graphicx}
\usepackage{amssymb}
\usepackage{esint}

\makeatletter
\@ifundefined{textcolor}{}
{%
 \definecolor{BLACK}{gray}{0}
 \definecolor{WHITE}{gray}{1}
 \definecolor{RED}{rgb}{1,0,0}
 \definecolor{GREEN}{rgb}{0,1,0}
 \definecolor{BLUE}{rgb}{0,0,1}
 \definecolor{CYAN}{cmyk}{1,0,0,0}
 \definecolor{MAGENTA}{cmyk}{0,1,0,0}
 \definecolor{YELLOW}{cmyk}{0,0,1,0}
 }

\@ifundefined{definecolor}
 {\usepackage{color}}{}
\makeatother

\makeatother

\usepackage{babel}

\makeatother

\usepackage{babel}

\makeatother

\usepackage{babel}

\begin{document}

\title{Probing anisotropic superfluidity of rashbons in atomic 
Fermi gases}

\author{Hui Hu$^{1}$, Lei Jiang$^{2}$, Xia-Ji Liu$^{1}$, and Han Pu$^{2}$}

\affiliation{$^{1}$ARC Centre of Excellence for Quantum-Atom Optics, Centre for
Atom Optics and Ultrafast Spectroscopy, Swinburne University of Technology,
Melbourne 3122, Australia \\
 $^{2}$Department of Physics and Astronomy, and Rice Quantum Institute,
Rice University, Houston, TX 77251, USA}

\date{\today}
\begin{abstract}
Motivated by the prospect of realizing a Fermi gas of $^{40}$K atoms
with a synthetic non-Abelian gauge field, we investigate theoretically
a strongly interacting Fermi gas in the presence of a Rashba spin-orbit
coupling. As the two-fold spin degeneracy is lifted by spin-orbit
interaction, bound pairs with mixed singlet and triplet pairings (referred
to as rashbons) emerge, leading to an anisotropic superfluid. We
show that this anisotropic superfluidity can be probed via measuring
the momentum distribution and single-particle spectral function in
a trapped atomic $^{40}$K cloud near a Feshbach resonance. 
\end{abstract}


\maketitle
Owing to the unprecedented experimental controllability, ultracold
atoms have been proven to be an ideal tabel-top system to study some
long-sought, challenging many-body problems. A well-known example
is the non-perturbative problem of crossover from a Bose-Einstein
condensation (BEC) to a Bardeen-Cooper-Schrieffer (BCS) superfluidity
in an ultracold atomic Fermi gases \cite{rmp}. Here
we study a strongly interacting Fermi gas in the presence
of a synthetic non-Abelian gauge field, as motivated by the recent
demonstration of such field in bosonic $^{87}$Rb atoms \cite{lin} and the prospect
of its realization in fermionic $^{40}$K atoms \cite{SOK}. We focus on the Rashba
spin-orbit (SO) interaction and explore its impact on the unitary Fermi gas.

A weak non-Abelian gauge field such as Rashba SO interaction is well
understood. It was shown in 2001 by Gor'kov and Rashba \cite{gor} that a condensed
matter system of superconducting 2D metals with weak SO coupling fetures a mixed spin singlet-triplet pairing field, and
its spin magnetic susceptibility can be dramatically affected by the
SO interaction. By applying an additional large Zeeman magnetic field,
it was proposed by Zhang {\em et al.} \cite{zhang} and Sato \textit{et al.} \cite{sato} that a topological
phase with gapless edge states and non-Abelian Majorana fermionic
quasiparticles may form. More recently, Vyasanakere and Shenoy identified
an interesting bound state by solving the two-body problem \cite{VS}, referred
to as \emph{rashbons}. By increasing the strength of SO coupling, a BCS
superfluid can therefore evolve into a BEC of rashbons \cite{VZS}. 

In this Letter, we investigate the salient features of rashbons by
using a functional path-integral functional method. We identify clearly
rashbons from the gaussian fluctuations of pairing field and show
they possess anistropic effective mass. We demonstrate at the saddle-point level
that the condensation of rashbons gives rise to both singlet and triplet
pairings. As a result, the momentum distribution and single-particle
spectral function become highly anisotropic.
By performing calculations for a realistic system of trapped $^{40}$K
fermions at a Feshbach resonance, we present observable experimental
signatures for visualizing the anisotropic superfluidity of rashbons.

\textit{The model --- }Let us start by formulating the BEC-BCS crossover
with a Rashba SO coupling $\mathcal{H}_{so}=\lambda(\hat{k}_{y}\hat{\sigma}_{x}-\hat{k}_{x}\hat{\sigma}_{y})$,
whose Hamiltonian is given by, \begin{equation}
{\cal H=}\int d{\bf r}\left\{ \psi^{+}\left[\xi_{{\bf k}}+\mathcal{H_{\textrm{so}}}\right]\psi+U_{0}\psi_{\uparrow}^{+}\psi_{\downarrow}^{+}\psi_{\downarrow}\psi_{\uparrow}\right\} ,\end{equation}
 where $\xi_{{\bf k}}=\hbar^{2}\hat{k}^{2}/(2m)-\mu$, and $\psi({\bf r)}=[\psi_{\uparrow}({\bf r)},\psi_{\downarrow}({\bf r)}]$
denotes collectively the fermionic field operators. The contact $s$-wave interaction
($U_{0}<0$) occurs between un-like spins. We use the functional path
integral method \cite{sademelo} and consider the action ${\cal Z}=\int{\cal D}[\psi,\bar{\psi}]\exp\left\{ -S\left[\psi\left({\bf r},\tau\right),\bar{\psi}\left({\bf r},\tau\right)\right]\right\} $,
where $S[\psi,\bar{\psi}]=\int_{0}^{\beta}d\tau[\int d{\bf r}\sum_{\sigma}\bar{\psi}_{\sigma}\left({\bf r}\right)\partial_{\tau}\psi_{\sigma}\left({\bf r}\right)]+{\cal H(}\psi,\bar{\psi})$,
$\beta=1/(k_{B}T)$, and ${\cal H}(\psi,\bar{\psi})$ is obtained
by replacing the field operators $\psi^{+}$ and $\psi$ with the
Grassmann variables $\bar{\psi}$ and $\psi$, respectively. The interaction
term can be decoupled by using the standard Hubbard-Stratonovich transformation
with the introduction of a fluctuating pairing field $\Delta({\bf r},\tau)$.
It is convenient to use the 4-dimensional Nambu spinor $\Phi\left({\bf r,}\tau\right)\equiv[\psi_{\uparrow},\psi_{\downarrow}{\bf ,}\bar{\psi}_{\uparrow},\bar{\psi}_{\downarrow}]$
and rewrite the action as, ${\cal Z}=\int{\cal D}[\Phi,\bar{\Phi}{\bf ;}\Delta,\bar{\Delta}]\exp\{-\int_{0}^{\beta}d\tau\int d{\bf r}\left[\frac{1}{2}\bar{\Phi}\left(-{\cal G}^{-1}\right)\Phi-\frac{\left|\Delta\right|^{2}}{U_{0}}\right]-\beta\sum_{{\bf k}}\xi_{{\bf k}}\}$,
where ${\cal G}$ is the single-particle Green function.
%\begin{equation}
%{\cal G}^{-1}=\left[\begin{array}{cc}
%-\partial_{\tau}-\xi_{{\bf k}}-\mathcal{H}_{so} & i\Delta\hat{\sigma}_{y}\\
%-i\bar{\Delta}\hat{\sigma}_{y} & -\partial_{\tau}+\xi_{{\bf k}}-\mathcal{H}_{so}^{*}\end{array}\right].\end{equation}
% We note that the Green function in the momentum space takes the following
%form, \begin{equation}
%{\cal G}\left({\bf k},i\omega_{m}\right)=\left[\begin{array}{cc}
%\hat{g}({\bf k},i\omega_{m}) & \hat{f}({\bf k},i\omega_{m})\\
%\hat{f}^{+}(\mathbf{k},-i\omega_{m}) & -\hat{g}^{T}(-{\bf k},-i\omega_{m})\end{array}\right],\end{equation}
% where the $2\times2$ matrix $\hat{g}({\bf k},i\omega_{m})$ and
%$\hat{f}({\bf k},i\omega_{m})$ satisfy the relations, $\hat{g}_{12}({\bf k},i\omega_{m})=\hat{g}_{21}^{*}({\bf k},-i\omega_{m})$
%and $\hat{f}_{12}({\bf k},i\omega_{m})=-\hat{f}_{21}(-{\bf k},-i\omega_{m})$.

By integrating out the fermionic field, the effective action is ${\cal Z}=\int{\cal D}[\Delta,\bar{\Delta}]\exp\left\{ -S_{eff}\left[\Delta,\bar{\Delta}\right]\right\} $,
where $S_{eff}=\int_{0}^{\beta}d\tau\int d{\bf r}\left\{ -\frac{\left|\Delta\left({\bf r},\tau\right)\right|^{2}}{U_{0}}\right\} -\frac{1}{2}$Tr$\ln\left[-{\cal G}^{-1}\right]+\beta\sum_{{\bf k}}\xi_{{\bf k}}$.
To proceed, we restrict ourselves to the gaussian fluctuation and expand $\Delta\left({\bf r},\tau\right)=\Delta_{0}+\delta\Delta\left({\bf r},\tau\right)$.
The effective action is then approximated by $S_{eff}=S_{0}+\Delta S$,
where the saddle-point action $S_{0}=\int_{0}^{\beta}d\tau\int d{\bf r}\left(-\frac{\Delta_{0}^{2}}{U_{0}}\right)-\frac{1}{2}$Tr$\ln\left[-{\cal G}_{0}^{-1}\right]+\beta\sum_{{\bf k}}\xi_{{\bf k}}$
and, in the momentum space, the fluctuation action takes the form
[$k\equiv\left({\bf k},i\omega_{m}\right)$ and $q\equiv\left({\bf q},i\nu_{n}\right)$]:
$\Delta S=\sum_{{\bf q},i\nu_{n}}\left[-\frac{1}{U_{0}}\delta\Delta(q)\delta\bar{\Delta}(q)\right]+\frac{1}{2}\left(\frac{1}{2}\right)$Tr$_{\sigma}\sum_{k,q}\left[{\cal G}_{0}\left(k\right)\Sigma\left(q\right){\cal G}_{0}\left(k-q\right)\Sigma\left(-q\right)\right]$,
where \begin{equation}
\Sigma\left(q\right)=\left[\begin{array}{cc}
0 & i\delta\Delta\left(q\right)\hat{\sigma}_{y}\\
-i\delta\bar{\Delta}\left(-q\right)\hat{\sigma}_{y} & 0\end{array}\right].\end{equation}

\textit{Rashbon --- }Let us consider first the normal state with
$\Delta_{0}=0$, in which case the Green function reduces to its non-interacting form as ${\cal G}_0(k) = {\rm Diag} \{ \hat{g}_0(k), -\hat{g}_0(-k) \}$ with
$\hat{g}_{0}(k)=[i\omega_{m}-\xi_{{\bf k}}-\lambda(k_{y}\hat{\sigma}_{x}-k_{x}\hat{\sigma}_{y})]^{-1}$,
leading to \emph{two} helicity branches in the single-particle spectrum,
$E_{{\bf k},\alpha}=\xi_{{\bf k}}+ \alpha \lambda k_{\perp}$, where $k_\perp \equiv ({k_x^2+k_y^2})^{1/2}$ and $\alpha = \pm 1$. The fluctuation action is given by $
\Delta S=\sum_{q}\left[-\Gamma^{-1}\left(q\right)\right]\delta\Delta(q)\delta\bar{\Delta}(q) $,
 where 
%$\Gamma^{-1}(q)=1/U_{0}+(1/2)k_{B}T\sum_{k}\{\sum_{\alpha}1/[(i\omega_{m}-E_{{\bf k},\alpha})(i\nu_{n}-i\omega_{m}-E_{{\bf q}-{\bf k},\alpha})]-A_{res}\}$ is the inverse vertex function,
%with $A_{res}=\lambda^{2}[k_{\perp}({\bf q}-{\bf k})_{\perp}+k_{x}(q_{x}-k_{x})+k_{y}(q_{y}-k_{y})]/[(i\omega_{m}-E_{{\bf k},+})(i\omega_{m}-E_{{\bf k},-})(i\nu_{n}-i\omega_{m}-E_{{\bf q}-{\bf k},+})(i\nu_{n}-i\omega_{m}-E_{{\bf q}-{\bf k},-})]$.
$\Gamma^{-1}(q)$ is the inverse vertex function which, at ${\bf q=0}$, takes the form \begin{equation}
\Gamma^{-1}\left(\omega\right)=\frac{m}{4\pi\hbar^{2}a_{s}}-\frac{1}{V}\sum_{{\bf k}}\left[\sum_{\alpha=\pm}\frac{1/2-f\left(E_{{\bf k},\alpha}\right)}{\omega^{+}-2E_{{\bf k},\alpha}}+\frac{1}{2\epsilon_{{\bf k}}}\right],\nonumber \end{equation}
 where $f\left(x\right)=1/(e^{x/k_{B}T}+1)$ is the Fermi distribution
function and we have renormalized the bare interaction $U_{0}$ by
the \textit{s}-wave scattering length, $1/U_{0}=m/(4\pi\hbar^{2}a_{s})-V^{-1}\sum_{{\bf k}}1/(2\epsilon_{{\bf k}})$, with $V$ being the quantization volume.

\begin{figure}[htp]

\begin{centering}
\includegraphics[clip,width=0.4\textwidth]{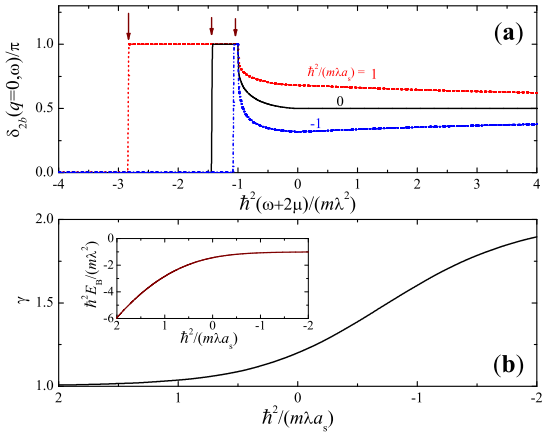} 
\par\end{centering}

\caption{(Color online) (a) Rashbons as evidenced by the two-body phase shift
of $\Gamma^{-1}\left({\bf 0},\omega\right)$ at three different scattering
lengths. The arrows indicate the position of binding energy. (b) Effective
mass of rashbons [$\gamma = M_\perp/(2m)$] in the strong SO limit. The inset
shows the bound state energy as a function of the scattering length. }

\label{fig1} 
\end{figure}

The vertex function is simply the Green function of the fermion pair.
A bound state can therefore be examined clearly by calculating the
phase shift \cite{nsr} $\delta({\bf q},\omega)=-\mathop{\rm Im}\ln[-\Gamma^{-1}({\bf q},i\nu_{n}\rightarrow\omega+i0^{+})]$.
For a true boson, the phase shift is given by $\delta_{B}({\bf q},\omega)=\pi\Theta(\omega-\epsilon_{{\bf q}}^{B}+\mu_{B})$,
where $\epsilon_{{\bf q}}^{B}$ and $\mu_{B}$ are the bosonic dispersion
and chemical potential, respectively, and $\Theta(x)$ is the step
function. In Fig.~\ref{fig1}(a), we plot the two-body part of the phase shift
at ${\bf q=0}$, obtained by discarding Fermi functions. The phase shift
jumps from $0$ to $\pi$ at a critical frequency, resembling
that of a true boson. This is exactly the demonstration of a bound
state, or rashbon. By recalling that the bosonic chemical potential
is given by $\mu_{B}=2\mu-E_{B}$, where $E_{B}$ is the bound state energy,
the critical frequency $(\omega+2\mu)_{c}$ at ${\bf q=0}$ gives
exactly $E_{B}$. Using the fact that the critical frequency corresponds to the position
where ${\rm Re}[\Gamma^{-1}]$ changes sign, we have
\begin{equation}
\frac{m}{4\pi\hbar^{2}a_{s}}-\frac{1}{2V}\sum_{{\bf k};\alpha=\pm}\left[\frac{1}{E_{B}-2E_{{\bf k},\alpha}}+\frac{1}{\epsilon_{{\bf k}}}\right]=0\,.\end{equation}
The inset in Fig.~\ref{fig1}(b) shows the bound state energy as a function of the
SO coupling strength. At the unitarity limit, the bound state energy is
universally given by $E_{B}(a_{s}=\pm\infty)\approx-1.439229m\lambda^{2}/\hbar^{2}$.
The size of rashbons $a$ is therefore at the order of $\hbar^{2}/(m\lambda)$.
Rashbons are well defined once $a\ll k_{F}^{-1}$ or $\lambda k_{F}\gg\epsilon_{F}$.
Thus, we anticipate that the system will cross over to a gas of rashbons
at $\lambda k_{F}/\epsilon_{F}\sim1$.

In the limit of a large SO coupling, the well-defined rashbons should
have a bosonic dispersion $\epsilon_{{\bf q}}^{B}=\hbar^{2}q_{\perp}^{2}/(2M_{\perp})+\hbar^{2}q_{z}^{2}/(2M_{z})$
and weakly interact with each other \emph{repulsively}. Because of
the anisotropic fermionic dispersion $E_{{\bf k},\pm}=\xi_{{\bf k}}\pm\lambda k_{\perp}$,
the effective mass of rashbons becomes anisotropic. While $M_{z}=2m$ is
not affected by the Rashba coupling, $M_{\perp}$ may get strongly
renormalized. As the jump of the phase shift at nonzero ${\bf q}$ which
occurs at $(\omega+2\mu)_{c}-\epsilon_{{\bf q}}^{B}$, we can numerically
determine $M_{\perp}$. Fig.~\ref{fig1}(b) reports $\gamma=M_{\perp}/(2m)$.
At unitarity, we find $\gamma \simeq 1.2$. When the system becomes an ensemble of weakly interacting rashbons, 
the heavy mass $M_{\perp}$ causes a decrease in the condensation
temperature so that $T_{\rm BEC}=\gamma^{-2/3}T_{\rm BEC,0}$,
where $T_{\rm BEC,0}\simeq 0.218T_{F}$ is the BEC temperature without the SO coupling.

%At intermediate SO coupling, rashbons coexist with fermions and interact
%very strongly. The transition temperature can be determined by taking
%into accout the number of rashbons, which can be calculated from the
%vertex function. In the unitarity limit, we anticipate that $T_{c}$
%is insensitive to SO coupling, since the critical temeprature in the
%absence of SO coupling, $T_{c}\simeq 0.15T_{F}$, is very close to
%the critical temperature in the strong SO coupling limit, $ $$T_{c}=\gamma^{-2/3}T_{\rm BEC,0}\simeq 0.16T_{F}$. 

\begin{figure}[htp]
\begin{centering}
\includegraphics[clip,width=0.48\textwidth]{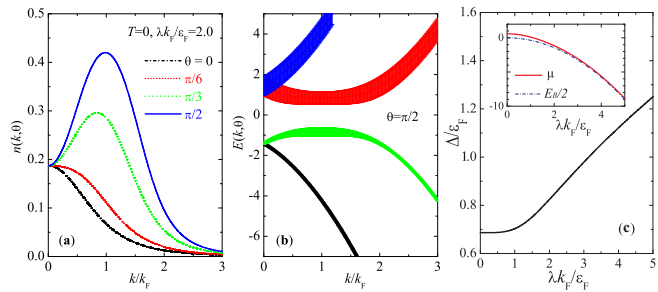} 
\par\end{centering}

%\caption{(a) Mean-field order parameter as a function of the SO coupling for
%a homogeneous unitary Fermi gas at zero temperature. The inset shows
%the chemical potential and the half of binding energy. (b) Temperature
%dependence of the order parameter at different SO coupling strengths.}

\caption{(a) Momentum distribution and (b) single-particle spectral function for $\theta=\pi/2$ at $\lambda k_F/\epsilon_F=2$. Here $\theta$ is the angle between ${\bf k}$ and the $z$-axis. The width of the curves in (c) represents the weight factor $(1\pm \gamma_{{\bf k},\pm})/4$ for each of the four Bogoliubov excitations. (c) Mean-field order parameter as a function of the SO coupling for
a homogeneous unitary Fermi gas at zero temperature. The inset shows
the chemical potential and the half of bound state energy, in units of $\epsilon_F$.}

\label{fig2} 
\end{figure}

\textit{Condensation of rashbons --- }Let us now turn to the condensed
phase characterized by a nonzero order parameter $\Delta_{0}\neq 0$.
At the mean-field saddle-point level, the single-particle Green function
takes the form, \begin{equation}
{\cal G}_{0}^{-1}=\left[\begin{array}{cc}
i\omega_{m}-\xi_{{\bf k}}-\mathcal{H}_{so} & i\Delta_{0}\hat{\sigma}_{y}\\
-i\Delta_{0}\hat{\sigma}_{y} & i\omega_{m}+\xi_{{\bf k}}-\mathcal{H}_{so}^{*}\end{array}\right].\label{gf0}\end{equation}
 The inversion of the above matrix can be worked out explicitly, leading to \emph{two}
single-particle Bogoliubov dispersions whose degeneracy is lifted by the SO interaction,
$E_{{\bf k},\pm}=\sqrt{\left(\xi_{{\bf k}}\pm\lambda k_{\perp}\right)^{2}+\Delta_{0}^{2}}$, and the normal and anomalous Green functions from which we can immediately obtain 
%Defining $\gamma_{{\bf k},\pm}=\left(\xi_{{\bf k}}\pm \lambda k_{\perp}\right)/E_{{\bf k},\pm}$,
%the normal and anomalous Green functions are \begin{eqnarray*}
%\hat{g}_{11} & = &\hat{g}_{22}= \frac{1}{4}\sum_{\alpha=\pm}\left[\frac{1+\gamma_{{\bf k},\alpha}}{i\omega_{m}-E_{{\bf k},\alpha}}+\frac{1-\gamma_{{\bf k},\alpha}}{i\omega_{m}+E_{{\bf k},\alpha}}\right],\\
%\hat{g}_{12} & = & \frac{1}{4}\sum_{\alpha=\pm}\alpha e^{i\varphi_{{\bf k}}}\left[\frac{1+\gamma_{{\bf k},\alpha}}{i\omega_{m}-E_{{\bf k},\alpha}}+\frac{1-\gamma_{{\bf k},\alpha}}{i\omega_{m}+E_{{\bf k},\alpha}}\right],\\
%\hat{f}_{11} & = &-\hat{f}^*_{22}= \sum_{\alpha=\pm}\frac{\alpha e^{-i\varphi_{{\bf k}}}}{4}\left[\frac{\Delta_{0}/E_{{\bf k},\alpha}}{i\omega_{m}-E_{{\bf k},\alpha}}-\frac{\Delta_{0}/E_{{\bf k},\alpha}}{i\omega_{m}+E_{{\bf k},\alpha}}\right],\\
%\hat{f}_{12} & = & \frac{1}{4}\sum_{\alpha=\pm}\left[\frac{\Delta_{0}/E_{{\bf k},\alpha}}{i\omega_{m}+E_{{\bf k},\alpha}}-\frac{\Delta_{0}/E_{{\bf k},\alpha}}{i\omega_{m}-E_{{\bf k},\alpha}}\right],\end{eqnarray*}
% where $e^{i\varphi_{{\bf k}}}\equiv(k_{y}+ik_{x})/k_{\perp}$. 
%We obtain immediately 
the momentum distribution $n\left({\bf k}\right)=1-\sum_{\alpha}\left[1/2-f\left(E_{{\bf k},\alpha}\right)\right]\gamma_{{\bf k},\alpha}$
and the single-particle spectral function $A_\uparrow({\bf k},\omega)=A_\downarrow({\bf k},\omega)=\sum_{\alpha}\left[\left(1+\gamma_{{\bf k},\alpha}\right)\delta\left(\omega-E_{{\bf k},\alpha}\right)+\left(1-\gamma_{{\bf k},\alpha}\right)\delta\left(\omega+E_{{\bf k},\alpha}\right)\right]/4$, where $\gamma_{{\bf k},\pm}=\left(\xi_{{\bf k}}\pm \lambda k_{\perp}\right)/E_{{\bf k},\pm}$.
The chemical potential and the order parameter are to be determined by
the number and the  gap equations, $n=\sum_{{\bf k}}n ({\bf k} )$
and $\Delta_{0}=-U_{0}\Delta_{0}\sum_{\alpha}[1/2-f\left(E_{{\bf k},\alpha}\right)]/(2E_{{\bf k},\alpha})$,
respectively. Fig.~\ref{fig2} displays the chemical potential and order parameter
as functions of the SO coupling strength for a unitary Fermi gas.
The increase of the SO strength leads to a deeper bound state. In analogy
with the BEC-BCS crossover, the order parameter and critical transition
temperature are greatly enhanced once the rashbons are well defined
at $\lambda k_{F}\sim\epsilon_{F}$. In the large SO coupling limit,
the chemical potential can be written as $\mu=E_{B}/2+\mu_{B}/2$,
where the chemical potential of rashbons $\mu_{B}$ is positive due
to the repulsion between rashbons and decreases with increasing coupling
as shown in the inset of Fig.~\ref{fig2}(c). By assuming an \textit{s}-wave repulsion
with scattering length $a_{B}$, where $\mu_{B}\simeq(n/2)4\pi\hbar^{2}a_{B}/M$,
we estimate within mean-field that in the unitarity limit, $a_{B}\simeq 3\hbar^{2}/(m\lambda)$,
comparable to the size of rashbons. A more accurate description of the
interaction between two rashbons can be obtained by solving a four-fermion
problem.

Figure~\ref{fig2}(a) and (b) illustrate the momentum distribution and the single-particle spectral function, respectively. These quantities exhibit anisotropic distribution in momentum space due to the SO coupling and can be readily measured in experiment.

\begin{figure}[htp]
\begin{centering}
\includegraphics[clip,width=0.4\textwidth]{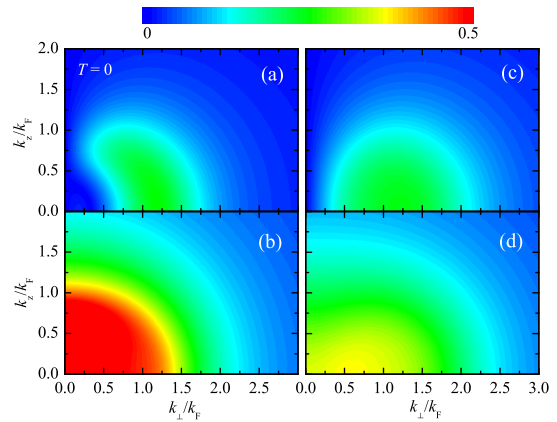} 
\par\end{centering}

\caption{(Color online) Linear contour plot for the triple pairing correlation $|\left\langle \psi_{{\bf k}\uparrow}\psi_{-{\bf k}\uparrow}\right\rangle|$
between like spins (upper panel) and the singlet pairing correlation $|\left\langle \psi_{{\bf k}\uparrow}\psi_{-{\bf k}\downarrow}\right\rangle |$
between un-like spins (lower panel) for a homogeneous unitary Fermi
gas at zero temperature. The SO coupling strength in (a) and (b) is
$\lambda k_{F}/\epsilon_{F}=1$, and in (c) and (d) is $\lambda k_{F}/\epsilon_{F}=2$.}

\label{fig3} 
\end{figure}

Another important consequence of the SO coupling is that the pairing field contains both a singlet and a triplet component \cite{gor,VZS}. For the system under study, it is straightforward to show that the triplet and singlet pairing fields are given by $\left\langle \psi_{{\bf k}\uparrow}\psi_{-{\bf k}\uparrow}\right\rangle =-i\Delta_{0}e^{-i\varphi_{{\bf k}}}\sum_{\alpha}\alpha[1/2-f\left(E_{{\bf k},\alpha}\right)]/(2E_{{\bf k},\alpha})$ and $\left\langle \psi_{{\bf k}\uparrow}\psi_{-{\bf k}\downarrow}\right\rangle =\Delta_{0}\sum_{\alpha}[1/2-f\left(E_{{\bf k},\alpha}\right)]/(2E_{{\bf k},\alpha})$, respectively, where $e^{-i\varphi_{{\bf k}}} \equiv (k_x-ik_y)/k_\perp$. The magnitude of these are shown in Fig.~\ref{fig3}.
The weight of the triplet component increases as the SO coupling strength increases and becomes equal to that of the singlet component in the strong SO coupling limit.

%
%\begin{figure}[htp]
%\begin{centering}
%\includegraphics[clip,width=0.4\textwidth]{fig4} 
%\par\end{centering}
%
%\caption{(Color online) Momentum distribution (a) and single-particle spectral
%function (b) of a homogeneous unitary Fermi gas at $\lambda k_{F}/\epsilon_{F}=2$.}
%
%
%\label{fig4} 
%\end{figure}

%
%The significant triplet pairing component of rashbons implies the
%anisotropy in the momentum distribution $n({\bf k})$ and single-particle
%spectral function $A({\bf k},\omega)$ in their condensed phase. In
%Fig. 4, we present $n(k_{\perp},k_{z}=k\cos\theta)$ at zero temperature
%and $A(k_{\perp},k_{z}=k\cos\theta;\omega)$ at a large SO coupling
%$\lambda k_{F}/\epsilon_{F}=2$. For the momentum distribution (Fig.
%4a), the anisotropy is evident in the strong dependence on angle $\theta$.
%While for the single-particle spectral function, one should be observe
%a splitting from two to four Bogoliubov excitations with increasing
%$\theta$ from $0$ to $\pi/2$. The error bars in Fig. 4b give the
%weight $\left(1\pm\gamma_{{\bf k},\pm}\right)/4$ for the four Bogoliubov
%excitations at $\theta=\pi/2$.

%
\begin{figure}[htp]
\begin{centering}
\includegraphics[clip,width=0.4\textwidth]{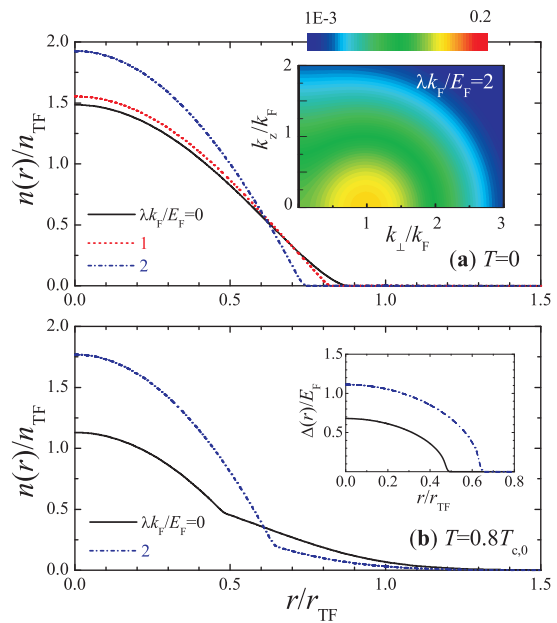} 
\par\end{centering}

\caption{(Color online) Density distributions of a trapped unitary Fermi gas
at zero $T=0$ (a) and finite temperature $T=0.8T_{c,0}$ (b) and at different SO coupling
strengths. $T_{c,0}$ is the critical temperature in the absence of the SO coupling. The inset in (a) shows log-scale contour plot of the momentum distribution and the inset in (b) illustrates the profile of the
order parameter at finite temperature. Here the Fermi energy is given by $E_{F}=(3N)^{1/3}\hbar\omega_{0}$,
Fermi wave number $k_{F}=(24N)^{1/6}a_{ho}^{-1}$, Thomas-Fermi radius
$r_{TF}=(24N)^{1/6}a_{ho}$, and the non-interacting peak density $n_{TF}=(24N)^{1/2}/(3\pi^{2})a_{ho}^{-3}$, where $a_{ho}=\sqrt{\hbar/(m\omega)}$ is the characteristic length
of the harmonic oscillator.}

\label{fig5} 
\end{figure}

\textit{Probing the anisotropic superfluidity of rashbons} --- One leading candidate to observe superfluid rashbons is a trapped
Fermi gas of $^{40}$K atoms near a broad Feshbach resonance, where
an applicable scheme to generate the Rashba SO coupling was recently
proposed \cite{SOK}. Previous experiments have demonstrated the measurement of momentum
distribution and single-particle spectral function is $^{40}K$ in the absence of the SO coupling \cite{jin}. We perform the
mean-field calculation in a 3D spherical harmonic trap $V_{T}(r)=m\omega_{0}^{2}r^{2}/2$,
by using the local density approximation (LDA). In LDA, the gas was divided
into small cells with a local chemical potential $\mu(r)=\mu-V_{T}(r)$.
The local density $n(r)$, momentum distribution $n({\bf k};r)/(2\pi)^{3}$,
occupied spectral function $A({\bf k},\omega;r)f\left(\omega\right)k^{2}/(2\pi^{2})$
are then integrated over the whole space to obtain the total contribution.
The chemical potential at the trap center $\mu$ can be determined
by the number equation $\int d{\bf r}n(r)=N$, where $N$ is the total
number of fermions. 
We show, in Fig.~\ref{fig5}, the density profile
of a trapped unitary Fermi gas at different SO coupling strengths
and temperatures. As anticipated, with the increase of the SO coupling
the cloud shrinks. The anisotropic momentum distribution at large SO coupling, which can be measured using the time-of-flight technique, can still be clearly seen as illustrated in the inset of Fig.~\ref{fig5}(a).

%
%\begin{figure}[htp]
%\begin{centering}
%\includegraphics[clip,width=0.4\textwidth]{fig6} 
%\par\end{centering}
%
%\caption{(Color online) Log-scale contour plot of the momentum distribution
%for a trapped unitary Fermi gas at different SO coupling strengths.}
%
%
%\label{fig6} 
%\end{figure}

%
\begin{figure}[htp]
\begin{centering}
\includegraphics[clip,width=0.4\textwidth]{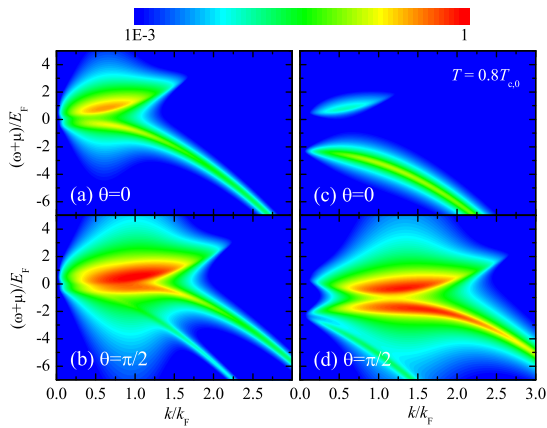} 
\par\end{centering}

\caption{(Color online) Log-scale contour plot of the single-particle spectral
function for a trapped unitary Fermi gas at $T=0.8T_{c,0}$. The SO coupling strength in (a) and (b) is
$\lambda k_{F}/E_{F}=1$, and in (c) and (d) is $\lambda k_{F}/E_{F}=2$.}

\label{fig7} 
\end{figure}

%Figure \ref{fig7} presents the contour plot of the momentum distribution for
%a trapped unitary Fermi gas at zero temperature. With increasing the
%SO coupling, the isotropic distribution (Fig. 6a) evolves gradually
%into a highly anisotropic shape with a maximum located at a nonzero
%$k_{\perp}$ (Fig. 6c). This evolution in anisotropy can be easily
%measured by absorption imaging, by suddenly switching off both the
%Feshbach magnetic field and the optical dipole trap for a gas of $^{40}$K
%atoms.

Figure \ref{fig7} presents the occupied spectral function at $T=0.8T_{c,0}$, where $T_{c,0}$ is the critical temperature
without the SO interaction. The distinct behavior for the spectral function
at $\theta=0$ (along the $z$-axis) and $\pi/2$ (in the transverse plane) can be probed by the recently demonstrated momentum-resolved
rf-spectroscopy \cite{jin}. To make better comparison with the experiment, we
have included in the calculation an energy resolution of $0.2E_{F}$,
as presented in the JILA rf-measurement. Our calculations therefore clearly demonstrate that the anisotropic nature of the rashbon superfluid will not be smeared out by averaging over the trapped cloud.

\textit{Summary} --- We have shown that rashbons induced by a strong
Rashba spin-orbit coupling differ significantly from the BEC-BCS crossover
molecules studied over the past few years. Rashbons have anisotrpic effective
mass and condense into a mixed singlet-triplet pairing state. They
lead to a strong anisotropy in the momentum distribution and the single-particle
spectral function. We have proposed that these distinct behaviors
can be readily probed in a strongly interacting Fermi gas of $^{40}$K
atoms in a synthetic non-Abelian gauge field.

More interesting features of rashbons may be discovered thanks to
the unprecedented controllability in ultracold atoms. Rashbons in
2D may be utilized to create Majorana fermions \cite{zhang,tewari}. Rashbons under a magnetic
field may exhibit inhomogeneous superfluidity with mixed singlet-triplet
pairings and their properties may be greatly affected by the field. Therefore, the exploration of strong correlation effects
of rashbons is a new exciting many-body problem. In the current work, we have adopted a mean-field approach. The mean-field calculation in the condensed phase can be improved by incorporating gaussian pair fluctuations \cite{hld}.

\textit{Acknowledgment} --- HH and XJL was supported by the ARC Discovery
Projects No. DP0984522 and No. DP0984637. HP is supported by the NSF and the Welch Foundation (Grant No. C-1669). We would like to thank Chuanwei Zhang, Hui Zhai and Shizhong Zhang for useful discussions.

{\bf Note added:} After we completed our work, we became aware of the work reported in Ref.~\cite{zhai} which treats a similar system. Our results agree with each other where they overlap.

\end{document}